\documentstyle[aps,prc,preprint]{revtex}
\tightenlines
\begin {document}
\draft
\preprint{SUNY-NTG-97-48}
\title
{A Microscopic Calculation of Photoabsorption Cross Sections on Protons 
and Nuclei}
 
\author
{R. Rapp$^{1}$, M. Urban$^2$, M. Buballa$^2$ and J. Wambach$^2$}
 
\address
{1) Department of Physics, State University of New York, 
    Stony Brook, NY 11794-3800, U.S.A.\\
 2) Institut f\"ur Kernphysik, TH Darmstadt, Schlo{\ss}gartenstr.9,
    D-64289 Darmstadt, Germany }
 
\maketitle
 
\begin{abstract}
A recently developed model for $\rho$-meson propagation in dense hadronic
matter is applied to total photoabsorption cross sections in 
$\gamma$-proton and $\gamma$-nucleus reactions. Within the vector dominance 
model the photon coupling to the virtual pion cloud of the nucleon, 
two-body meson-exchange currents, as well as $\gamma$-nucleon resonances are 
included. Whereas the $\gamma p$ reaction is determined by the 
low-density limit of the model, higher orders in the nuclear density
are important to correctly account for the experimental spectra
observed on both light and heavy nuclei over a wide range of
photon energies, including the region below the pion threshold.  
In connection with soft dilepton spectra in high-energy heavy-ion collisions
we emphasize the importance of photoabsorption to further constrain 
the parameters of the model.
\end{abstract}


Recent measurements of dilepton spectra in heavy-ion collisions at both 
intermediate and high bombarding energies have shown a strong enhancement
of the pair yield in the low-invariant-mass region 
($M$$\simeq$0.2-1.5~GeV) as compared to expectations based 
on free hadronic sources\cite{CERES,HELIOS3,DLS}. So far, the most 
promising approaches to explain these data are based on medium 
modifications in $\pi^+\pi^-\to l^+l^-$ annihilation 
occurring during the interaction phase of the 
colliding nuclear system. In particular, the assumption of a dropping 
$\rho$-meson mass~\cite{BR91} in the hot and dense medium has been 
shown to give
a good description of the HELIOS-3 and CERES data~\cite{LKBS,BrCa}. 
On the other hand, the inclusion of in-medium hadronic interactions in 
$\rho$- and $\pi\pi$ propagation also gives reasonable agreement with 
these experiments~\cite{RCW,CBRW}. It is evident that any model 
of dilepton enhancement has to be in accordance  
with a wide variety of related data, {\it e.g.} the free 
$\rho$ meson in the time-like region has to be accounted
for by properly describing p-wave $\pi\pi$
scattering and the pion electromagnetic form factor. 
Obviously another important constraint is provided by 
photoabsorption experiments~\cite{KlWe,SYZ}, which represent the limit
of vanishing invariant mass of the (virtual) photon, $M^2$$\to$0.
In this note we will extend our model for $\rho$-meson propagation in 
hadronic matter~\cite{RCW} to analyze photoabsorption 
spectra on both protons and nuclei, thereby further constraining 
the model parameters. 

The starting point is the general expression for the total absorption 
cross section of a photon on a volume element $d^3x$ of cold 
nuclear matter 
\begin{equation} 
\frac{d\sigma}{d^3x}=-\frac{4\pi\alpha}{q_0} \ \epsilon_\mu(q,\lambda)
\ \epsilon_\nu(q,\lambda) \ {\rm Im}G^{\mu\nu}(q_0,\vec{q};\rho_N) \ . 
\label{cross1} 
\end{equation}     
Here, $\epsilon_\mu$ and $(q_0,\vec q)$ denote the photon polarization 
vector and four-momentum, respectively, while $G_{\mu\nu}$ represents the 
electromagnetic current correlator of the hadronic source at a 
given nuclear density $\rho_N$. Invoking the vector dominance model (VDM)
and neglecting small contributions from isoscalar vector mesons  
the correlator is determined by the $\rho$-meson propagator as 
\begin{equation}   
G_{\mu\nu}(q_0,q;\rho_N) \equiv \frac{(m_\rho^{(0)})^4}{g^2} \ 
D_\rho^{\mu\nu}(q_0,q;\rho_N) 
\label{vdm} 
\end{equation}
where 
\begin{eqnarray} 
D_\rho^{\mu\nu}(q_0,\vec q ;\rho_N) & = & 
\frac{P_L^{\mu\nu}}{M^2-(m_\rho^{(0)})^2 -\Sigma_\rho^L(q_0,\vec q; \rho_N)}
+\frac{P_T^{\mu\nu}}{M^2-(m_\rho^{(0)})^2 -\Sigma_\rho^T(q_0,\vec q;\rho_N)}
\nonumber\\
 & & \qquad \qquad +\frac{q^\mu q^\nu}{(m_\rho^{(0)})^2 M^2} \ . 
\label{drhomunu}
\end{eqnarray}
Here $P_L^{\mu\nu}$ ($P_T^{\mu\nu}$) is the standard 
longitudinal (transverse) projection operator and
$\Sigma_\rho^L$ ($\Sigma_\rho^T$) 
the corresponding scalar part of the selfenergy. 
For real photons with $M^2=q_0^2-{\vec q}^2=0$ Eq.~(\ref{cross1}) can be 
rewritten as the total photoabsorption cross section normalized to the number
of nucleons, $A$, as  
\begin{equation}  
\frac{\sigma_{\gamma A}^{\rm abs}}{A}=-\frac{4\pi\alpha}{q_0} \ 
\frac{(m_\rho^{(0)})^4}{g^2} \frac {1}{\rho_N} \ 
{\rm Im}D_\rho^T(q_0,\vec q ;\rho_N) \ ,  
\label{cross2} 
\end{equation}
where $D_\rho^T$ denotes the transverse $\rho$-meson propagator, defined 
through $\Sigma^T_\rho$.\\  
For the actual calculations we have to specify a model for the in-medium 
$\rho$-meson propagator. It consists of a bare $\rho$ with 
mass $m_\rho^{(0)}$ renormalized through various selfenergy contributions  
involving two-pion and single-baryon interactions 
\begin{equation} 
\Sigma_\rho^{L,T} = \Sigma_{\rho\pi\pi}^{L,T} + \Sigma_{\rho N}^{L,T} \ .
\label{Siglt}
\end{equation}
In free space ($\rho_N$=0), only $\Sigma_{\rho\pi\pi}^0(M)\equiv
\Sigma_{\rho\pi\pi}(q_0,\vec q;\rho_N=0)$ survives, representing
the coupling of the $\rho$ to vacuum two-pion states. Aside from 
the two-pion loop we include a pion-tadpole contribution rendering
$\Sigma_{\rho\pi\pi}^0(M)$ transverse and zero at the photon point, once 
properly regularized \cite{Urban}. The parameters are fixed by ensuring a good
description of the p-wave $\pi\pi$ phase shifts as well as the pion 
electromagnetic form factor in vacuum. At finite density
$\Sigma_{\rho\pi\pi}$ 
is modified through pion interactions with the surrounding
nucleons~\cite{RCW}.
As is well known, the dominant contribution to the in-medium pion propagator 
arises from p-wave nucleon-hole and delta-hole polarizations. The
corresponding 
selfenergies contain coupling constants $f_{\pi\alpha}$ 
($\alpha=NN^{-1},\Delta N^{-1}$) related via
$f_{\pi N\Delta}$=2$f_{\pi NN}$ (Chew-Low factor~\cite{ChLo}), a monopole
form factor
\begin{equation}
F_{\pi\alpha}(k)= \left( \frac{\Lambda_\pi^2-m_\pi^2}{\Lambda_\pi^2
+k^2} \right) \ , 
\end{equation}
spin-isospin factors $SI(\pi\alpha)$ and a Lindhard function 
$\phi_\alpha(\omega,k)$ for the loop integration over the nucleon Fermi sea
(see below). The precise value of the cutoff parameter $\Lambda_\pi$ will 
be determined from the fit to the photoabsorption data.   
Furthermore, the pion selfenergies have to be corrected for
short-range correlation effects, conveniently parameterized in terms of
Migdal parameters $g'_{\alpha\beta}$. The calculations of 
ref.~\cite{RCW} were done in back-to-back kinematics ($\vec q=0$)
for simplicity. 
When going to the photon point, however, one necessarily has to allow 
for finite 3-momentum of the $\rho$ meson relative to the rest frame of
the nuclear medium. Recently this has been achieved by Urban et al.
maintaining exact conservation of the hadronic vector 
current and will be discussed separately~\cite{Urban}. 
In the low-density limit, the in-medium $\rho$-meson 
propagator per nucleon reduces to the forward Compton amplitude on the
proton. The model specified above describes the coupling of 
the photon to the virtual pion cloud of the nucleon via an intermediate 
$\rho$ meson and yields non-resonant 'background contributions'
to the Compton amplitude.\\
The second piece of the in-medium $\rho$ selfenergy, $\Sigma_{\rho N}$,
in Eq.~(\ref{Siglt}) 
arises from direct coupling of the $\rho$ meson to the surrounding 
nucleons leading to nucleonic resonances. 
We assume the $\rho N$ amplitude to be governed  by s-channel pole 
graphs. This was first discussed in ref.~\cite{FrPi} for the case 
of the $N(1720)$ and $\Delta(1905)$, which both show a large branching 
ratio ($>$60\%) into the $\rho N$ channel. However, 
the photoabsorption data (especially for the free proton) require the 
inclusion of additional, lower-lying resonances. We account for the most 
important states which will allow us to saturate the experimental 
spectra. They can be divided into two groups: 
\begin{itemize}  
\item[(i)] positive parity states, which exhibit a predominant $p$-wave 
coupling to $\rho N$. In the non-relativistic limit, suitable interaction 
Lagrangians are given by  
\begin{equation} 
{\cal L}_{\rho BN}^{p-wave}  =  \frac{f_{\rho BN}}{m_\rho} \
\Psi^\dagger_{B} \ ({\vec s} \times {\vec q}) \cdot
 \vec{\rho}_a \ t_a \ \Psi_N \ + \  h.c. \ ,  
\label{lpw} 
\end{equation}   
where the summation over $a$ is in isospin space. The spin operators 
$\vec{s}=\vec{\sigma}, \vec{S}$ for $J$=1/2,3/2 and the isospin operators 
$\vec{t}=\vec{\tau}, \vec{T}$ for $I$=1/2,3/2, respectively,  are chosen 
in accordance with the quantum numbers of baryon $B$ for $B$=$N(939)$, 
$\Delta(1232)$ and $N(1720)$, where $\sigma_j$ and $\tau_a$
are the usual Pauli matrices and $S_j$, $T_a$ the $J,I=1/2 \to 3/2$ transition
operators. For $B$=$\Delta(1905)$, which carries spin $J$=5/2, a tensor 
coupling of type $(R_{ij} \ q_i \ \rho_{j,a} \ T_a)$ is
employed~\cite{FrPi}.   
\item[(ii)]
negative  parity states, which exhibit a predominant $s$-wave
coupling to $\rho N$. In the non-relativistic limit, the interaction
Lagrangians can be chosen as~\cite{Lenske}  
\begin{equation}
{\cal L}_{\rho BN}^{s-wave}  =  \frac{f_{\rho BN}}{m_\rho} \
\Psi^\dagger_{B} \ (q_0 \ {\vec s}\cdot \vec{\rho}_a -  
\rho^0_a \ {\vec s}\cdot {\vec q}) \ t_a \ \Psi_N \ + \ h.c. \ 
\label{lsw} 
\end{equation}   
for $B$=$N(1520)$, $\Delta(1620)$, $\Delta(1700)$. 
\end{itemize} 
From these interaction vertices we derive in-medium 
selfenergy tensors for $\rho$-like $BN^{-1}$ excitations. In close analogy 
to the pionic case one obtains for the purely transverse $p$-wave 
contributions 
\begin{equation}
\Sigma_{\rho\alpha,pw}^{(0),T}(q_0,q) =
 -  \left(\frac{f_{\rho\alpha} \ F_{\rho\alpha}(q)}{m_\rho}
\right)^2 \ SI(\rho\alpha) \ q^2 \ \phi_{\rho\alpha}(q_0,q) \ , 
\end{equation} 
while for $s$-waves  
\begin{equation}
\Sigma_{\rho\alpha,sw}^{(0),T}(q_0,q)  =  
 - \left(\frac{f_{\rho\alpha} \ F_{\rho\alpha}(q)}{m_\rho}
\right)^2 \ SI(\rho\alpha) \ q_0^2 \ \phi_{\rho\alpha}(q_0,q) \ ,
\end{equation} 
with a monopole form factor
$F_{\rho\alpha}(q)=\Lambda_\rho^2/(\Lambda_\rho^2+q^2)$ 
and spin-isospin factors $SI(\rho\alpha)$ summarized in table~\ref{tab1}
(note that for $M^2 = 0$ we have $q^2 = q_0^2$ and thus the expressions 
for s-wave and p-wave coupling become identical).  
In analogy to pion-induced excitations we include short-range
correlation effects in the particle-hole bubble through Migdal parameters
$g'$, which also induce a mixing between the various excitations of a given 
partial wave~\cite{RCW}. The explicit form of the Lindhard functions reads  
\begin{equation}
\phi_{\rho\alpha}(q_0,\vec q)=-\int_0^{p_F} \frac{p^2 dp}{(2\pi)^2} 
\int\limits_{-1}^{+1}
 \ dx \ \sum_{+,-} \ \frac{1}
 {\pm q_0+E_p^N-E_{pq}^{B}(x) \pm
 \frac{i}{2}\Gamma_{B}^{tot}} \ ,  
\label{Lindhard}
\end{equation}
which is equivalent to the pionic case ($E_p^N$=$\sqrt{m_N^2+p^2}$, 
$E_{pq}^B(x)$=$\sqrt{m_B^2+q^2+p^2+2pqx}$ ).   
The free baryon widths are each taken as the sum of 
$\rho N$ channel and $\pi N$ channel in the appropriate partial wave, 
\begin{equation}
\Gamma_{B}^{0}(s)=\Gamma^0_{B\to \rho N}(s)+
\Gamma^0_{B\to \pi N}(s) \ .  
\label{gamma0}
\end{equation} 
In the nuclear medium we account for a density-dependent correction as 
\begin{equation}
\Gamma_{B}(s;\rho)=\Gamma_{B}^0(s)+\Gamma_{B}^{med} 
\ \frac{\rho}{\rho_0} \ ,
\end{equation}
where possible energy dependencies in $\Gamma_{B}^{med}$
have been neglected for simplicity. 
It remains to fix the 
coupling constants $f_{\rho BN}$. For B=$N(939)$ and $\Delta(1232)$ 
we take values close to the Bonn potential; all other resonances 
considered have a sizable branching ratio into the $\rho N$ channel. These
branching 
ratios are used to obtain an estimate for the $\rho BN$ coupling constants via
\begin{eqnarray}
\Gamma^0_{B\to N\rho}(\sqrt{s}) & = & \frac{f_{\rho BN}^2}{4\pi m_\rho^2} 
\ \frac{2m_N}{\sqrt{s}} \ \frac{(2I_\rho+1)}{(2J_B+1)(2I_B+1)} \
SI(\rho B N) 
\nonumber\\ 
 & &  \qquad \qquad  \int\limits_{2m_\pi}^{\sqrt{s}-m_N} 
\frac{M dM}{\pi} A^0_\rho(M) \ q_{cm} \ F_\rho(q_{cm})^2 \ f(M) \ , 
\label{gammaNrho0} 
\end{eqnarray}
where the kinematic factor $f(M)$ is given by $f(M)=q_{cm}^2$ for p-wave 
coupling and $f(M)=(2q_0^2+M^2)$ for s-wave coupling with 
\begin{equation}
q_{cm}^2=\frac{(s-M^2-m_N^2)^2-4m_N^2M^2}{4s} \ 
\end{equation}
being the $\rho$/$N$ decay momentum in the resonance rest frame.  
$A_\rho^0(M)=-2 \ {\rm Im} D^0_\rho(M)$ denotes the $\rho$-meson spectral 
function in the vacuum. 
The $\pi BN$ coupling constant used for $\Gamma_{B\to\pi N}^0$ in 
Eq.~(\ref{gamma0}) is then chosen such 
that the total width matches its experimental value at the resonance mass
$s=m_B^2$. \\
As has been noted long ago the most simple version of the VDM, 
Eq.~(\ref{vdm}), typically results in an overestimation of the 
$B\to N\gamma$ branching fractions when using the hadronic coupling 
constants deduced from the $B\to N\rho$ partial widths. However, one  
can correct for this by employing an improved version of the VDM~\cite{KLZ},
which allows to adjust the $BN\gamma$ coupling $\mu_B$ 
(the transition magnetic moment) at the photon 
point independently~\cite{FrPi}. It amounts to replacing the combination 
$(m_\rho^{(0)})^4\;{\rm Im}D_\rho^T(q_0,\vec q ;\rho_N)$ 
entering Eq.~(\ref{cross2}) by the following 'transition form factor':
\begin{eqnarray} 
\bar{F}(q_0,q;\rho_N) & = & -{\rm Im}\Sigma^T_{\rho\pi\pi} |d_\rho-1|^2 
-{\rm Im}\Sigma^T_{\rho N} |d_\rho-r_B|^2 
\nonumber\\
d_\rho(q_0,q;\rho_N) & = & 
\frac{ M^2-\Sigma^T_{\rho\pi\pi}- r_B \Sigma^T_{\rho N} }  
{ M^2-(m_\rho^{(0)})^2-\Sigma^T_{\rho\pi\pi}-\Sigma^T_{\rho N} }
\end{eqnarray} 
where 
\begin{equation} 
r_B=\frac{\mu_B}{\frac{f_{\rho BN}}{m_\rho} \ 
\frac{(m_\rho^{(0)})^2}{g}} 
\end{equation} 
denotes the ratio of the photon coupling to its value in the naive VDM. 
In principle each resonance state $B$ can be assigned a separate 
value for $r_B$ but, as will be seen below, reasonable fits to the 
photoabsorption spectra can be achieved with a common value for both 
$r_B$ and $\Lambda_\rho$, making use of some freedom in the hadronic
couplings $f_{\rho BN}$  within the experimental uncertainties of the 
partial widths, Eq.~(\ref{gammaNrho0}).
The final formula to be used for the photoabsorption calculations then 
reads 
\begin{equation}
\frac{\sigma_{\gamma A}^{\rm abs}}{A}=-\frac{4\pi\alpha}{g^2 q_0} \
 \frac {1}{\rho_N} \ \bar{F}(q_0,q ;\rho_N) \ .
\label{cross3}
\end{equation}
In taking the low-density limit, $\rho_N\to 0$, only terms linear 
in density contribute to $\bar{F}$, representing the absorption process 
on a single nucleon, as mentioned above. Fig.~1 shows the resulting cross
section 
with the free parameters taken as $\Lambda_\pi$=550~MeV,
$\Lambda_\rho$=600~MeV
($\Lambda_{\rho NN}$=1~GeV)  and $r_B$=0.5 (the actual hadronic coupling
constants 
$f_{\rho BN}$ are given in table~\ref{tab1}). The 
various contributions from $\rho$-meson coupling to
the virtual pion cloud of the nucleon (the 
'background' described by $\Sigma_{\rho\pi\pi}$) as well as from the 
$\rho N$ resonances (contained in $\Sigma_{\rho N}$) add incoherently.\\
For absorption spectra on nuclei, 
one experimentally observes an almost independent scaling with the 
atomic number $A$ of different nuclei ({\it cp.}~Fig.~2). 
This justifies to perform the calculations 
in infinite nuclear matter at an average 
density, which we take to be $\bar{\rho}_N$=0.8$\rho_0$ (we have checked that 
the results for the normalized cross section, Eq.~(\ref{cross3}), 
only weakly depend on density within reasonable limits). 
As compared to the free proton two additional features appear in the nuclear
medium: short-range correlation effects in the resummation of 
the particle-hole bubbles and in-medium 
corrections to the resonance widths. Due to the rather soft 
form factors involved, the $p$-wave excitations turn out to favor 
rather small Landau-Migdal parameters of   
$g'_{NN}$=0.6 and $g'_{\alpha\beta}$=0.25 for all other transitions, 
whereas the $s$-wave bubbles show no significant evidence
for short-range correlations (therefore we set $g'_{s-wave}\equiv$0).    
Note that the rather large in-medium correction to the $N$(1520) width, 
$\Gamma_{N(1520)}^{med}$=250~MeV, can be understood microscopically 
in a selfconsistent treatment of the $\rho$ spectral function
in nuclear matter~\cite{Lenske}. On the other hand, the net in-medium
correction to the $\Delta$(1232) width is quite small. This
reflects the fact that a moderate in-medium
broadening is largely compensated by Pauli blocking effects on the 
decay nucleon. The sensitivity of our results with respect to the in-medium
widths of the higher lying resonances is comparatively small. As can be 
seen from
Fig.~2, a reasonable fit is obtained up to incident photon energies of
about 1~GeV.  Beyond that the inclusion of further baryon resonances 
in both the $\pi N$ and $\rho N$ interactions as well as higher partial 
waves seems to be required. It is noteworthy that below the pion
threshold some strength appears. This is nothing but the 
well-known 'quasi deuteron' tail above the giant dipole resonance, arising
from pion-exchange currents. These are naturally included in our model.     

In summary, we have shown that our earlier model for rho-meson 
propagation in hadronic matter~\cite{RCW} allows for a consistent 
application at the photon point ($M^2$=0).  
With additional improvements on both the two-pion selfenergy 
$\Sigma_{\rho\pi\pi}$ (including the full 3-momentum dependence) and 
resonant $\rho N$ contributions (including $s$-wave contributions
as well as an improved version of the VDM) an acceptable description 
of total photoabsorption cross sections on both the proton and nuclei 
has been achieved with a rather limited number of parameters, 
thereby further constraining their actual values. This clearly increases
the confidence in the model when applying it to calculate dilepton 
production as  measured in relativistic heavy-ion collisions
at various bombarding energies. Indeed, employing our model in a transport 
theoretical  analysis of the CERN experiments (CERES and HELIOS-3) gives 
good agreement with the observed dilepton spectra~\cite{CBRW}.

\vskip1cm
 
\centerline {\bf ACKNOWLEDGMENTS}
We are grateful to J. Ahrens  for providing us with the data compilation  
of the experimental photoabsorption cross sections on nuclei. We
thank B. Friman, F. Klingl, A. Richter and W. Weise for fruitful
discussions.     
One of us (RR) acknowledges support 
from the Alexander-von-Humboldt foundation as a Feodor-Lynen fellow. 
This work is supported in part by the U.S. Department of Energy 
under Grant No. DE-FG02-88ER40388.

\newpage

\begin{table}
\caption{\it Properties of the $\rho BN$ vertices as derived from the 
interaction lagrangians, Eqs.~(7) and~(8); table columns from left to 
right: baryon resonance, relative angular momentum in the $\rho N$ decay, 
spin-isospin factor (note that in its definition we have absorbed an 
additional factor of $\frac{1}{2}$ as compared to table 2 in 
ref.~[7]), 
partial decay width into $\rho N$ as extracted from ref.~[16], 
coupling constant as estimated from $\Gamma^0_{\rho N}$ (for N(939) and 
$\Delta$(1232) we have indicated the values from the BONN 
potential~[17] which uses  
somewhat harder form factors), coupling constant as actually used in our
photoabsorption fit, in-medium correction to the total decay width. }
\label{tab1}
\begin{tabular}{ccccccc}
 B & $l_{\rho N}$ & $SI(\rho BN^{-1})$ & $\Gamma^0_{\rho N}$ [MeV] &  
$\left(\frac{f_{\rho BN}^2}{4\pi}\right)_{est}$ & 
$\left(\frac{f_{\rho BN}^2}{4\pi}\right)_{fit}$ & 
$\Gamma^{med}$ [MeV] \\
\hline
N(939)         & $p$ & 4    & --   & 4.68  & 5.8  & 0  \\
$\Delta$(1232) & $p$ & 16/9 & --   & 18.72 & 23.2 & 15   \\
$N$(1520)      & $s$ &  8/3 & 24   & 6.95  & 5.5  & 250 \\
$\Delta$(1620) & $s$ &  8/3 & 22.5 & 1.01  & 0.7  & 50  \\
$\Delta$(1700) & $s$ & 16/9 & 45   & 1.2   & 1.2  & 50  \\
$N$(1720)      & $p$ &  8/3 & 105  & 8.99  & 9.2  & 50  \\
$\Delta$(1905) & $p$ &  4/5 & 210  & 17.6  & 18.5 & 50  \\
\end{tabular}
\end{table}
%
%
%
%

\pagebreak

%
%
\begin{center}
{\large \sl \bf Figure Captions}
\end{center}
\vspace{0.5cm}

\begin{itemize}
\item[{\bf Figure 1}:]  
Total photoabsorption cross section on the proton: full result of
our fit (solid line), $\pi\pi$ 'background' (dashed line) as well as 
the three dominant $\rho N$ resonances $\Delta$(1232), $N$(1520)  
and $N$(1720) (dashed-dotted lines). The data are taken from 
ref.~\cite{Arm}.

\item[{\bf Figure 2:}]  
Total photoabsorption cross sections on different nuclei: the solid
line represents our full result while the dashed line denotes the 
contribution of the $\pi\pi$ background only, both calculated at a nuclear 
density $\bar{\rho}_N$=0.8$\rho_0$. The data are compiled from 
refs.~\cite{Saclay,Ahrens,Mainz,Frascati}.  

\end{itemize}

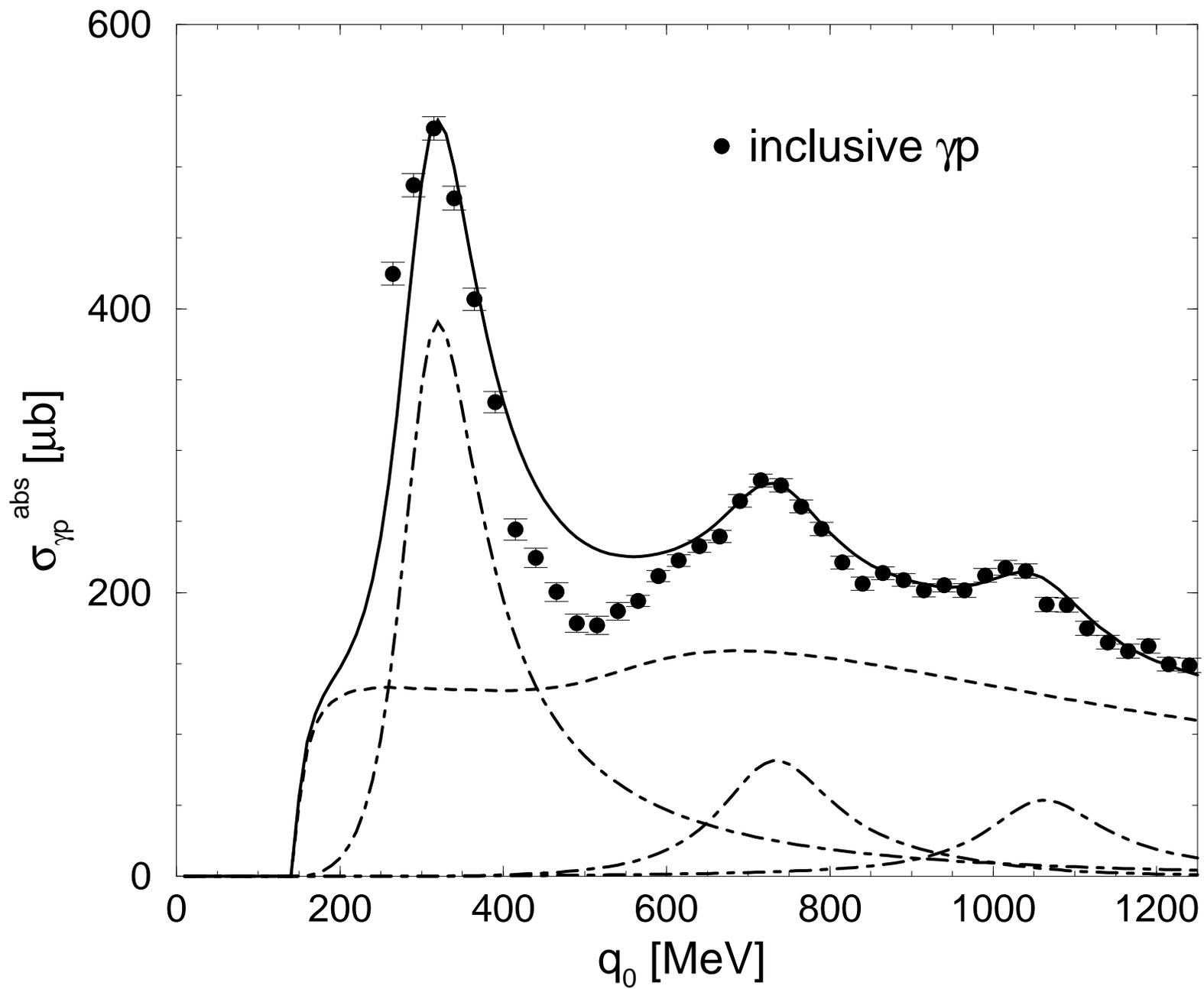
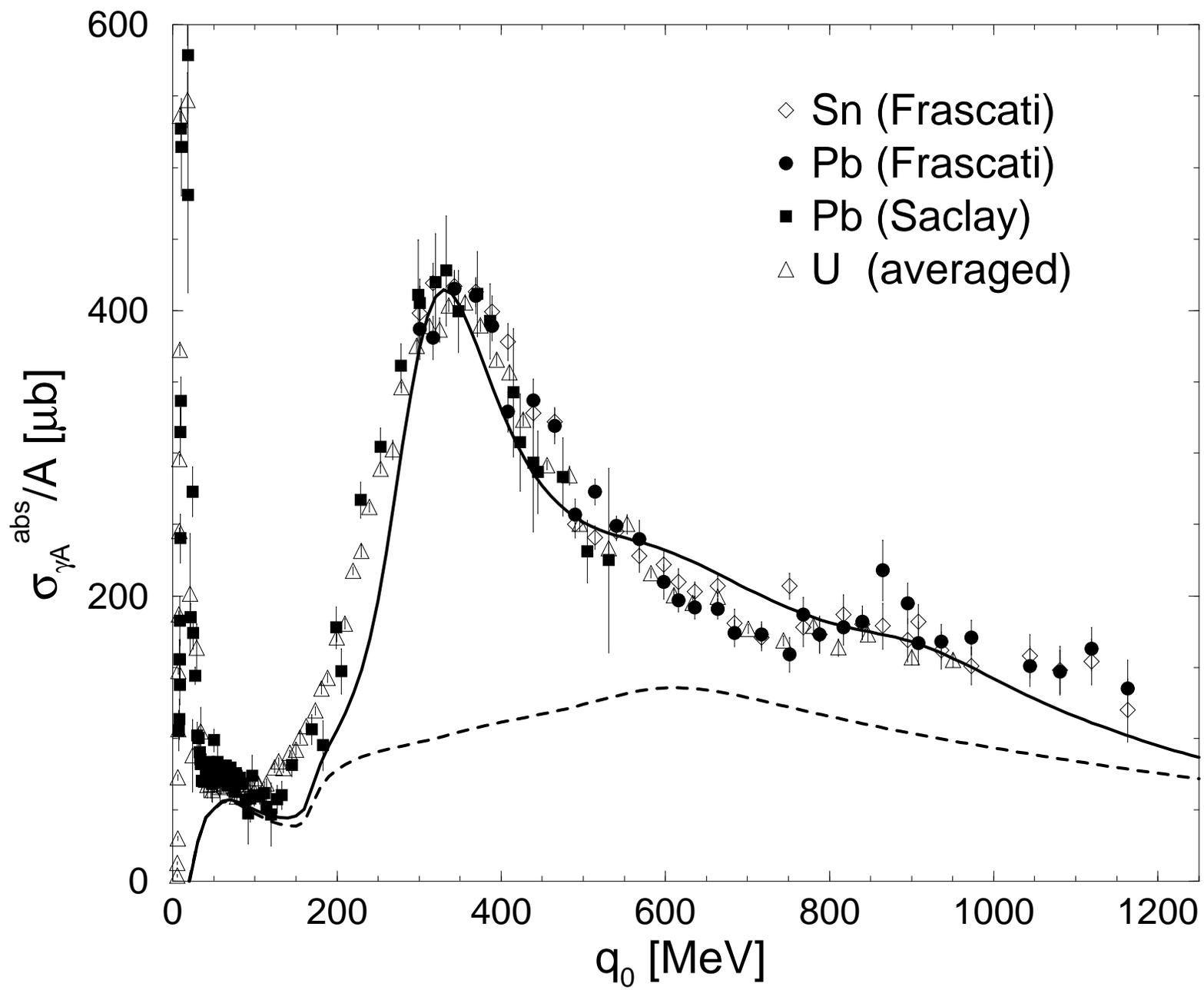
\end{document}